\documentclass[sigconf]{acmart}
\usepackage{balance}
\usepackage{booktabs} 
\usepackage{adjustbox}
\usepackage{multirow}
\usepackage{multicol}

\usepackage{bbm}
\usepackage{amsmath,bm}
\usepackage{subfig}
\usepackage{tikz}
\usetikzlibrary{arrows.meta}
\usetikzlibrary{positioning, fit, mindmap, trees, calc,tikzmark,shapes}
\usetikzlibrary{shapes.arrows, fadings, automata,tikzmark,decorations.pathreplacing,patterns}

\usepackage{graphicx}

\usepackage{sistyle}
\SIthousandsep{,}

\usepackage{xcolor}

\definecolor{quartz}{RGB}{219,223,238}
\definecolor{spring_sun}{RGB}{242,243,195}
\definecolor{dairy_cream}{RGB}{254,226,189}
\definecolor{surf_crest}{RGB}{205,230,208}
\definecolor{french_pass}{RGB}{195,232,246}
\definecolor{cosmos}{RGB}{248,209,210}
\definecolor{portafino}{RGB}{245,237,160}
\definecolor{sail}{RGB}{163,205,235}
\definecolor{hint_green}{RGB}{226,246,209}
\definecolor{bittersweet}{RGB}{255,111,105}
\definecolor{java}{RGB}{2,190,196}
\definecolor{ice_cold}{RGB}{169,232,220}
\definecolor{bgc}{RGB}{245,245,245}
\definecolor{tuatara}{RGB}{67, 67, 67}
\definecolor{aluminum}{RGB}{153,153,153}
\definecolor{silver}{RGB}{191,191,191}
\definecolor{platinum}{RGB}{228,228,228}
\definecolor{mercury}{RGB}{230,230,230}
\definecolor{gallery}{RGB}{240,240,240}
\definecolor{free_speech_aquamarine}{RGB}{0, 156, 114}
\definecolor{sun_shade}{RGB}{255, 144, 68}
\definecolor{fern}{RGB}{101,197,117}
\definecolor{french_blue}{RGB}{0, 112, 182}
\definecolor{matisse}{RGB}{25, 104, 167}
\definecolor{sushi}{RGB}{117, 168, 47}
\definecolor{shakespeare}{RGB}{85, 154, 193}
\definecolor{egg_shell}{RGB}{238, 234, 215}
\definecolor{carnation}{RGB}{245, 80, 86}
\definecolor{flamingo}{RGB}{237, 88, 85}
\definecolor{jet_stream}{RGB}{188, 214, 210}
\definecolor{jelly_bean}{RGB}{45, 126, 150}
\definecolor{tree_poppy}{RGB}{246, 154, 27}
\definecolor{deep_carmine_pink}{RGB}{236, 50, 67}
\definecolor{copper_rust}{RGB}{155, 64, 74}
\definecolor{midnight}{RGB}{0, 29, 50}
\definecolor{chilean_fire}{RGB}{215, 87, 44}
\definecolor{puerto_rico}{RGB}{94, 194, 166}
\definecolor{japanese_laurel}{RGB}{53, 116, 40}
\definecolor{fire_engine_red}{RGB}{206, 37, 51}
\definecolor{ku_crimson}{RGB}{243, 0, 25}
\definecolor{turmeric}{RGB}{211, 178, 76}
\definecolor{tahiti_gold}{RGB}{223, 102, 36}
\definecolor{outrageous_orange}{RGB}{255, 100, 45}
\definecolor{crusta}{RGB}{254, 127, 44}
\definecolor{safety_orange}{RGB}{254, 106, 0}
\definecolor{pigment_green}{RGB}{0, 175, 79}
\definecolor{jaffa}{RGB}{240, 131, 58}
\definecolor{jet_stream}{rgb}{0.69,0.61,0.85}
\definecolor{jelly_bean}{rgb}{0.47,0.32,0.66}
\definecolor{azalea}{RGB}{251, 196, 196}
\definecolor{sundown}{RGB}{249, 180, 181}
\definecolor{light_coral}{RGB}{244, 127, 123}
\definecolor{wewak}{RGB}{244, 143, 150}

\definecolor{biscay}{RGB}{44, 62, 80}
\definecolor{carmine_pink}{RGB}{231, 76, 60}
\definecolor{athens_gray}{RGB}{236, 240, 241}
\definecolor{celestial_blue}{RGB}{52, 152, 219}
\definecolor{curious_blue}{RGB}{41, 128, 185}
\definecolor{my_sin}{RGB}{255, 176, 59}
\definecolor{viridian}{RGB}{70, 137, 102}
\definecolor{tomato}{RGB}{255, 97, 56}
\definecolor{mountain_meadow}{RGB}{0, 163, 136}
\definecolor{padua}{RGB}{121, 189, 143}
\definecolor{killarney}{RGB}{56, 113, 66}
\definecolor{ocean_green}{RGB}{79, 176, 112}
\definecolor{pastel_green}{RGB}{107, 227, 135}
\definecolor{chinook}{RGB}{163, 232, 178}
\definecolor{cosmic_latte}{RGB}{222, 247, 229}
\definecolor{chateau_green}{RGB}{69, 191, 85}
\definecolor{RoyalBlue}{RGB}{69, 191, 85}

\definecolor{blue0}{RGB}{240,249,232}
\definecolor{blue1}{RGB}{204,235,197}
\definecolor{blue2}{RGB}{168,221,181}
\definecolor{blue3}{RGB}{123,204,196}
\definecolor{blue4}{RGB}{78,179,211}
\definecolor{blue5}{RGB}{43,140,190}
\definecolor{blue6}{RGB}{8,88,158}

\definecolor{yellow0}{RGB}{255,255,212}
\definecolor{yellow1}{RGB}{254,227,145}
\definecolor{yellow2}{RGB}{254,196,79}
\definecolor{yellow3}{RGB}{254,153,41}
\definecolor{yellow4}{RGB}{236,112,20}
\definecolor{yellow5}{RGB}{204,76,2}
\definecolor{yellow6}{RGB}{140,45,4}

\setcopyright{acmcopyright}

\begin{document}

\copyrightyear{2019} 
\acmYear{2019} 
\acmConference[RecSys '19]{Thirteenth ACM Conference on Recommender Systems}{September 16--20, 2019}{Copenhagen, Denmark}
\acmPrice{15.00}
\acmDOI{10.1145/3298689.3347000}
\acmISBN{978-1-4503-6243-6/19/09}

\title{Personalized Re-ranking for Recommendation }

\author{Changhua Pei$^{1\ast}$, Yi Zhang$^{1\ast}$, Yongfeng Zhang$^{2}$}
\authornote{Changhua Pei and Yi Zhang contribute equally. Yongfeng Zhang is the corresponding author.}
\author{Fei Sun$^1$, Xiao Lin$^1$,  Hanxiao Sun$^1$, Jian Wu$^1$, Peng Jiang$^3$,
Junfeng Ge$^1$, Wenwu Ou$^1$}
\affiliation{
\institution{$^1$ Alibaba Group\,\, $^2$ Rutgers University\,\, $^3$ Kwai Inc.}
{$^1$ \{changhua.pch, zhanyuan.zy, ofey.sf, hc.lx, hansel.shx, joshuawu.wujian, beili.gjf, santong.oww\}@alibaba-inc.com} \\
{$^2$ {yongfeng.zhang@rutgers.edu}\,\, $^3$ {jiangpeng@kuaishou.com}}
}

\renewcommand{\shortauthors}{Changhua Pei et al.}
\renewcommand{\arraystretch}{1.0}
\setlength{\textfloatsep}{5pt plus 2pt minus 2pt}
\setlength{\abovecaptionskip}{4pt}
\setlength{\belowcaptionskip}{2pt}
\begin{abstract}
Ranking is a core task in recommender systems, which aims at providing an ordered list of items to users. Typically, a ranking function is learned from the labeled dataset to optimize the global performance, which produces a ranking score for each individual item. However, it may be sub-optimal because the scoring function applies to each item individually and does not explicitly consider the mutual influence between items, as well as the differences of users' preferences or intents.
Therefore, we propose a personalized re-ranking model for recommender systems. The proposed re-ranking model can be easily deployed as a follow-up modular after any ranking algorithm, by directly using the existing ranking feature vectors. It directly optimizes the whole recommendation list by employing a transformer structure to efficiently encode the information of all items in the list. Specifically, the Transformer applies a self-attention mechanism that directly models the global relationships between any pair of items in the whole list. We confirm that the performance can be further improved by introducing pre-trained embedding to learn personalized encoding functions for different users.
Experimental results on both offline benchmarks and real-world online e-commerce systems demonstrate the significant improvements of the proposed re-ranking model.
\end{abstract}

%
%



\maketitle

\section{Introduction}
Ranking is crucial in recommender systems. The quality of the ranked list given by a ranking algorithm has a great impact on users' satisfaction as well as the revenue of the recommender systems. A large amount of ranking algorithms \cite{friedman2001greedy,burges2005learning,joachims2006training,burges2010ranknet,cao2007learning,taylor2008softrank,xia2008listwise} have been proposed to optimize the ranking performance. Typically ranking in recommender system only considers the user-item pair features, without considering the influences from other items in the list, especially by those items placed alongside\cite{carbonell1998use,zhai2015beyond}. Though \textit{pairwise} and \textit{listwise} learning to rank methods try to solve the problem by taking the item-pair or item-list as input, they only focus on optimizing the loss function to make better use of the labels, \textit{e.g.,} click-through data. They didn't explicitly model the mutual influences between items in the feature space.

Some works\cite{Yin:2016:RRY:2939672.2939677,zhuang2018globally,ai2018learning} tend to model the mutual influences between items explicitly to refine the initial list given by the previous ranking algorithm, which is known as re-ranking. The main idea is to build the scoring function by encoding intra-item patterns into feature space. The state-of-the-art methods for encoding the feature vectors are RNN-based, such as GlobalRerank\cite{zhuang2018globally} and DLCM\cite{ai2018learning}. They feed the initial list into RNN-based structure sequentially and output the encoded vector at each time step. However, RNN-based approaches have limited ability to model the interactions between items in the list. The feature information of the previous encoded item degrades along with the encoding distance. Inspired by the Transformer architecture\cite{kang2018self} used in machine translation, we propose to use the Transformer to model the mutual influences between items. The Transformer structure uses self-attention mechanism where any two items can interact with each other directly without degradation over the encoding distance. Meanwhile, 
the encoding procedure of Transformer is more efficient than RNN-based approach because of parallelization.

Besides the interactions between items, personalized encoding function of the interactions should also be considered for re-ranking in recommender system. Re-ranking for recommender system is user-specific, depending on the user's preferences and intents. For a user who is sensitive to price, the interaction between ``price'' feature should be more important in the re-ranking model. Typical global encoding function may be not optimal as it ignores the differences between feature distributions for each user. For instance, when users are focusing on price comparison, similar items with different prices tend to be more aggregated in the list. When the user has no obvious purchasing intention, items in the recommendation list tend to be more diverse. 
Therefore, we introduce a personalization module into the Transformer structure to represent user's preference and intent on item interactions. The interaction between items in the list and user can be captured simultaneously in our personalized re-ranking model.
  

The main contributions of this paper are as follows:
\begin{itemize}
	\item \textbf{Problem}. We propose a personalized re-ranking problem in recommender systems, which, to the best of our knowledge, is the first time to explicitly introduce the personalized information into re-ranking task in large-scale online system. The experimental results demonstrate the effectiveness of introducing users' representation into list representation for re-ranking.   
	\item \textbf{Model}. We employ the Transformer equipped with personalized embedding to compute representations of initial input ranking list and output the re-ranking score. The self-attention mechanism enable us to model user-specific mutual influences between any two items in a more effective and efficient way compared with RNN-based approaches.  
	\item \textbf{Data}. We release a large scale dataset (E-commerce Re-ranking dataset) used in this paper. This dataset is built from a real-world E-commerce recommender system. Records in the dataset contain a recommendation list for user with click-through labels and features for ranking.   
	\item \textbf{Evaluation}. We conducted both offline and online experiments which show that our methods significantly outperform the state-of-the-art approaches. The online A/B tests show that our approach achieves higher click-through rate and more revenue for real-world system.   
\end{itemize}

\section{Related Work}
Our work aims to refine the initial ranking list given by the base ranker. Among these base rankers, learning to rank is one of the widely used methods. The learning to rank methods can be classified into three categories according to the loss function they used: \textit{pointwise}\cite{cossock2008statistical,li2008mcrank}, \textit{pairwise}\cite{joachims2002optimizing,joachims2006training, burges2007learning}, and \textit{listwise}\cite{joachims2006training,burges2010ranknet, duan2010empirical,cao2007learning, xia2008listwise,xu2007adarank,taylor2008softrank}. All these methods learn a global scoring function within which the weight of a certain feature is globally learned. However, the weights of the features should be able to be aware of the interactions not only between items but also between the user and items. 

Closest to our work are \cite{ai2018learning,ai2018learningv2,zhuang2018globally,bello2018seq2slate}, which are all re-ranking methods. They use the whole initial list as input and model the complex dependencies between items in different ways. \cite{ai2018learning} uses unidirectional GRU\cite{cho2014properties} to encode the information of the whole list into the representation of each item. \cite{zhuang2018globally} uses LSTM\cite{Hochreiter:LSTM:NC1997} and \cite{bello2018seq2slate} uses pointer network\cite{vinyals2015pointer} not only to encode the whole list information, but also to generate the ranked list by a decoder. For those methods which use either GRU or LSTM to encode the dependencies of items, the capacity of the encoder is limited by the encoding distance. In our paper, we use transformer-like encoder, based on self-attention mechanism to model the interactions for any of two items in $\mathbf{O(1)}$ distance. Besides, for those methods which use decoder to sequentially generate the ordered list, they are not suitable for online ranking system which requires strict latency criterion. As the sequential decoder uses the item selected at time $t$-$1$ as input to select the item at time $t$, it can not be parallelized and needs $n$ times of inferences, where n is the length of the output list. \cite{ai2018learningv2} proposes a \textit{groupwise} scoring function which can be parallelized when scoring the items, but its computation cost is high  because it enumerates every possible combinations of items in the list. 
\section{Re-ranking Model Formulation}
\label{sec:formulation}
In this section, we first give some preliminary knowledge about learning to rank and re-ranking methods for recommendation systems. Then we formulate the problem we aim to solve in this paper. The notations used in this paper are in Table~\ref{tab:notation}.
 
\begin{table}[t]
\caption{Notation used in this paper.}
\centering
\begin{tabular}{lp{6cm}}
\toprule
 Notation. & Description. \\
\midrule
$\bm{X}$ & The matrix of features. \\
$\bm{PV}$ & The matrix of personalized vectors. \\
$\bm{PE}$ & The matrix of position embeddings. \\
$\bm{E}$ & The output matrix of the input layer.\\
$\mathcal{R}$ & The set of total users' requests. \\
$\mathcal{I}_r$ & The set of candidate items for each user's request $r \in \mathcal{R}$. \\
$\mathcal{S}_r$ & The initial list of items generated by the ranking approaches for each user's request $r$. \\
$\mathcal{H}_u$ & The sequence of items clicked by user $u$. \\
$\theta, \hat{\theta}, \theta^{'}$ & The parameter matrices of ranking, re-ranking and pre-trained model respectively. \\
$y_i$ & The label of click on item $i$. \\
$P(y_i|\cdot)$ & The click probability of item $i$ predicted by the model. \\ 
\bottomrule
\end{tabular}
\label{tab:notation}
\end{table}

Learning to rank (often labelled as \textbf{LTR}) method is widely used for ranking in real-work systems to generate an ordered list for information retrieval\cite{joachims2002optimizing,liu2009learning} and recommendation\cite{duan2010empirical}. The LTR method learns a global scoring function based on the feature vector of  items. Having this global function, the LTR method outputs an ordered list by scoring each item in the candidate set. This global scoring function is usually learned by minimizing the following loss function $\mathcal{L}$:  
\begin{equation}
	\mathcal{L} = \sum_{r \in \mathcal{R} }\ell \Bigl(\{ y_{i}, P(y_{i}|\bm{x_i};\theta)| i \in \mathcal{I}_r\}\Bigr)
	\label{eq:ltr-loss}
\end{equation}
where $\mathcal{R}$ is the set of all users' requests for recommendation. $\mathcal{I}_r$ is the candidate set of items for request $r \in \mathcal{R}$. $\bm{x_i}$ represents the feature space of item $i$. $y_{i}$ is the label on item $i$, \textit{i.e.,} click or not. $P(y_{i}|\bm{x_i};\theta)$ is the predicted click probability of item $i$ given by the ranking model with parameters $\theta$. $\ell$ is the loss computed with $y_{i}$ and $P(y_{i}|\bm{x_i};\theta)$.

However, $\bm{x_i}$ is not enough to learn a good scoring function. We find that ranking for recommender system should consider the following extra information: (a)  mutual influences between item-pairs\cite{carbonell1998use,zhai2015beyond}; (b) interactions between the users and items. The mutual influences between item-pairs can be directly learned from the initial list $\mathcal{S}_r = [i_1,i_2,...,i_n]$ given by the existing LTR model for the request $r$. Works\cite{ai2018learning}\cite{zhuang2018globally}\cite{ai2018learningv2}\cite{bello2018seq2slate} propose approaches to make better use of mutual information of item-pairs. However, few works consider the interactions between the users and items. The extent of mutual influences of item-pairs varies from user to user. In this paper, we introduce a personalized matrix $\bm{PV}$ to learn user-specific encoding function which is able to model personalized mutual influences between item-pairs. The loss function of the model can be formulated as Equation~\ref{eq:rr-loss}.     
\begin{equation}
	\mathcal{L} = \sum_{r \in \mathcal{R}}\ell\Bigl(\{ y_i, P(y_i|\bm{X}, \bm{PV};\hat{\theta})| i \in \mathcal{S}_r\}\Bigr)
	\label{eq:rr-loss}
\end{equation}
where $\mathcal{S}_r$ is the initial list given by the previous ranking model. $\hat{\theta}$ is the parameters of our re-ranking model. $\bm{X}$ is the feature matrix of all items in the list.



\section{Personalized Re-ranking Model}

In this section, we first give an overview of our proposed \textbf{P}ersonalized \textbf{R}e-ranking \textbf{M}odel (PRM). Then we introduce each component of our model in detail. 

\subsection{Model Architecture}
The architecture of PRM model is shown in Figure~\ref{fig:model1}. The model consists of three parts: the \textit{input} layer, the \textit{encoding} layer and the \textit{output} layer. It takes the initial list of items generated by previous ranking method as input and outputs a re-ranked list. 
The detailed structure will be introduced separately in the following sections.

\tikzset{
  ffn/.style = {draw, rectangle, fill=yellow3, minimum width=3em, minimum height=1.5em},
  item/.style = {draw, rectangle, fill=blue1, minimum width=3em, minimum height=1.5em},
  smallitem/.style = {draw, rectangle, fill=blue0, minimum width=2.8em, minimum height=1.8em},
  pvector/.style = {draw, rectangle, fill=blue2, minimum width=2.8em, minimum height=1.8em},
  transformer/.style = {draw, rectangle, fill=padua, minimum width=3em, minimum height=1.5em},
  norm/.style = {draw, rectangle, fill=blue1, minimum width=5.5em, minimum height=1.2em, inner sep=0, outer sep=0},
  dropout/.style = {draw, rectangle, fill=blue2, minimum width=4.5em, minimum height=1.2em, inner sep=0, outer sep=0},
  mh/.style = {draw, rectangle, fill=blue3, minimum width=4.5em, minimum height=1.5em},
  ff/.style = {draw, rectangle, fill=blue3, minimum width=4.5em, minimum height=1.5em},
  softmax/.style = {circle, draw, line width=0.5pt, minimum size=1em, inner sep=0pt, align=center},
  relu/.style = {draw, rectangle, fill=yellow3, minimum width=3.5cm, minimum height=0.5cm},
  relu2/.style = {draw, rectangle, fill=yellow2, minimum width=2.5cm, minimum height=0.5cm},
  sigmoid/.style = {draw, rectangle, fill=yellow1, minimum width=1.5cm, minimum height=0.5cm},
  pretrain_item1/.style = {draw, rectangle, fill=yellow4, minimum width=1.5cm, minimum height=0.5cm},
  pretrain_item2/.style = {draw, rectangle, fill=yellow3, minimum width=1.5cm, minimum height=0.5cm},
  pretrain_item3/.style = {draw, rectangle, fill=yellow2, minimum width=1.5cm, minimum height=0.5cm},
}

%
%


\begin{figure*}[htb]
\centering
\resizebox{0.99\textwidth}{!}{
\begin{tikzpicture}[]

 \pgfdeclarelayer{bg}    
\pgfsetlayers{bg,main}  
\def\vdistance{1.2cm}
\node[inner sep=0pt] (in) at (-2.8,0)
    {\includegraphics[width=18px,height=18px]{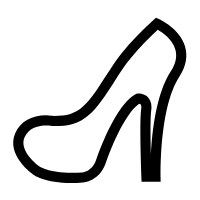}};
\node[above of=in, node distance=0.8cm] (i0) {$\vdots$};
\node[inner sep=0pt, above of=i0, node distance=0.6cm] (i3) {\includegraphics[width=18px,height=18px]{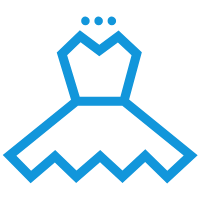}};

\node[inner sep=0pt, above of=i3, node distance=\vdistance] (i2) {\includegraphics[width=18px,height=18px]{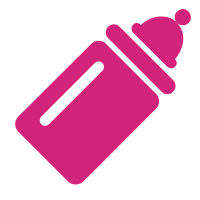}};

\node[inner sep=0pt, above of=i2, node distance=\vdistance] (i1) {\includegraphics[width=18px,height=18px]{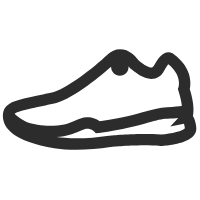}};

\node[smallitem, right of=in, node distance=1.2cm] (mn) {$\bm{x_{i_n}}$};
\node[right of=i0, node distance=1.2cm] (m0) {$\vdots$};
\node[smallitem, right of=i3, node distance=1.2cm] (m3) {$\bm{x_{i_3}}$};
\node[smallitem, right of=i2, node distance=1.2cm] (m2) {$\bm{x_{i_2}}$};
\node[smallitem, right of=i1, node distance=1.2cm] (m1) {$\bm{x_{i_1}}$};

\node[pvector, right of=mn, node distance=2.8em] (pvn) {$\bm{pv_{i_n}}$};
\node[right of=m0, node distance=2.8em] (pv0) {$\vdots$};
\node[pvector, right of=m3, node distance=2.8em] (pv3) {$\bm{pv_{i_3}}$};
\node[pvector, right of=m2, node distance=2.8em] (pv2) {$\bm{pv_{i_2}}$};
\node[pvector, right of=m1, node distance=2.8em] (pv1) {$\bm{pv_{i_1}}$};

 \draw[thick,dotted] ($(mn.south east)+(0.0, -0.3)$) rectangle ($(pv1.north east)+(0.0, 0.1)$);
\def\dpv_plus{1.0cm}
\node[softmax, right of=pvn, node distance=\dpv_plus] (plusn) {$+$};
\node[right of=pv0, node distance=\dpv_plus] (plus0) {};
\node[softmax, right of=pv3, node distance=\dpv_plus] (plus3) {$+$};
\node[softmax, right of=pv2, node distance=\dpv_plus] (plus2) {$+$};
\node[softmax, right of=pv1, node distance=\dpv_plus] (plus1) {$+$};

\def\pe_plus{0.6cm}
\node[below of=plusn, node distance=\pe_plus] (pen) {$\bm{pe_{i_n}}$};
\node[below of=plus3, node distance=\pe_plus] (pe3) {$\bm{pe_{i_3}}$};
\node[below of=plus2, node distance=\pe_plus] (pe2) {$\bm{pe_{i_2}}$};
\node[below of=plus1, node distance=\pe_plus] (pe1) {$\bm{pe_{i_1}}$};

\node[item, right of=pvn, node distance=2cm] (en) {$\bm{e_{i_n}}$};
\node[right of=pv0, node distance=2cm] (e0) {$\vdots$};
\node[item, right of=pv3, node distance=2cm] (e3) {$\bm{e_{i_3}}$};
\node[item, right of=pv2, node distance=2cm] (e2) {$\bm{e_{i_2}}$};
\node[item, right of=pv1, node distance=2cm] (e1) {$\bm{e_{i_1}}$};


\draw[-{Latex[length=1mm, width=0.8mm]}] ($(pvn.east)$) -- ($(plusn.west)$) ;
\draw[-{Latex[length=1mm, width=0.8mm]}] ($(pv3.east)$) -- ($(plus3.west)$) ;
\draw[-{Latex[length=1mm, width=0.8mm]}] ($(pv2.east)$) -- ($(plus2.west)$) ;
\draw[-{Latex[length=1mm, width=0.8mm]}] ($(pv1.east)$) -- ($(plus1.west)$) ;

\draw[-{Latex[length=1mm, width=0.8mm]}] ($(plusn.east)$) -- ($(en.west)$) ;
\draw[-{Latex[length=1mm, width=0.8mm]}] ($(plus3.east)$) -- ($(e3.west)$) ;
\draw[-{Latex[length=1mm, width=0.8mm]}] ($(plus2.east)$) -- ($(e2.west)$) ;
\draw[-{Latex[length=1mm, width=0.8mm]}] ($(plus1.east)$) -- ($(e1.west)$) ;

\draw[-{Latex[length=1mm, width=0.8mm]}] ($(pen.north)$) -- ($(plusn.south)$) ;
\draw[-{Latex[length=1mm, width=0.8mm]}] ($(pe3.north)$) -- ($(plus3.south)$) ;
\draw[-{Latex[length=1mm, width=0.8mm]}] ($(pe2.north)$) -- ($(plus2.south)$) ;
\draw[-{Latex[length=1mm, width=0.8mm]}] ($(pe1.north)$) -- ($(plus1.south)$) ;

\node[softmax, right of=en, node distance=2.4cm] (attn) {};
\node[right of=e0, node distance=2.4cm] (sf0) {$\vdots$};
\node[softmax, right of=e3, node distance=2.4cm] (att3) {};
\node[softmax, right of=e2, node distance=2.4cm] (att2) {};
\node[softmax, right of=e1, node distance=2.4cm] (att1) {};

\draw[-{Latex[length=1mm, width=0.8mm]}] ($(in.east)$) -- ($(mn.west)$) ;
\draw[-{Latex[length=1mm, width=0.8mm]}] ($(i3.east)$) -- ($(m3.west)$) ;
\draw[-{Latex[length=1mm, width=0.8mm]}] ($(i2.east)$) -- ($(m2.west)$) ;
\draw[-{Latex[length=1mm, width=0.8mm]}] ($(i1.east)$) -- ($(m1.west)$) ;

\draw[-{Latex[length=1mm, width=0.8mm]}] ($(en.east)$) -- ($(att3.west)$) ;
\draw[-{Latex[length=1mm, width=0.8mm]}] ($(en.east)$) -- ($(att2.west)$) ;
\draw[-{Latex[length=1mm, width=0.8mm]}] ($(en.east)$) -- ($(att1.west)$) ;
\draw[-{Latex[length=1mm, width=0.8mm]}] ($(en.east)$) -- ($(attn.west)$) ;

\draw[-{Latex[length=1mm, width=0.8mm]}] ($(e3.east)$) -- ($(att1.west)$) ;
\draw[-{Latex[length=1mm, width=0.8mm]}] ($(e3.east)$) -- ($(att2.west)$) ;
\draw[-{Latex[length=1mm, width=0.8mm]}] ($(e3.east)$) -- ($(att3.west)$) ;
\draw[-{Latex[length=1mm, width=0.8mm]}] ($(e3.east)$) -- ($(attn.west)$) ;

\draw[-{Latex[length=1mm, width=0.8mm]}] ($(e2.east)$) -- ($(att1.west)$) ;
\draw[-{Latex[length=1mm, width=0.8mm]}] ($(e2.east)$) -- ($(att2.west)$) ;
\draw[-{Latex[length=1mm, width=0.8mm]}] ($(e2.east)$) -- ($(att3.west)$) ;
\draw[-{Latex[length=1mm, width=0.8mm]}] ($(e2.east)$) -- ($(attn.west)$) ;
\draw[-{Latex[length=1mm, width=0.8mm]}] ($(e1.east)$) -- ($(att1.west)$) ;
\draw[-{Latex[length=1mm, width=0.8mm]}] ($(e1.east)$) -- ($(att2.west)$) ;
\draw[-{Latex[length=1mm, width=0.8mm]}] ($(e1.east)$) -- ($(att3.west)$) ;
\draw[-{Latex[length=1mm, width=0.8mm]}] ($(e1.east)$) -- ($(attn.west)$) ;

\node[ffn, right of=attn, node distance=1.4cm] (ffn_n) {FFN};
\node[right of=sf0, node distance=1.4cm] (ffn0) {$\vdots$};
\node[ffn, right of=att3, node distance=1.4cm] (ffn3) {FFN};
\node[ffn, right of=att2, node distance=1.4cm] (ffn2) {FFN};
\node[ffn, right of=att1, node distance=1.4cm] (ffn1) {FFN};

\draw[-{Latex[length=1mm, width=0.8mm]}] ($(attn.east)$) -- ($(ffn_n.west)$) ;
\draw[-{Latex[length=1mm, width=0.8mm]}] ($(att3.east)$) -- ($(ffn3.west)$) ;
\draw[-{Latex[length=1mm, width=0.8mm]}] ($(att2.east)$) -- ($(ffn2.west)$) ;
\draw[-{Latex[length=1mm, width=0.8mm]}] ($(att1.east)$) -- ($(ffn1.west)$) ;

\node[softmax, right of=ffn_n, node distance=1.4cm] (sfn) {};
\node[right of=ffn0, node distance=1.4cm] (sf0) {$\vdots$};
\node[softmax, right of=ffn3, node distance=1.4cm] (sf3) {};
\node[softmax, right of=ffn2, node distance=1.4cm] (sf2) {};
\node[softmax, right of=ffn1, node distance=1.4cm] (sf1) {};

\draw[-{Latex[length=1mm, width=0.8mm]}] ($(ffn_n.east)$) -- ($(sfn.west)$) ;
\draw[-{Latex[length=1mm, width=0.8mm]}] ($(ffn3.east)$) -- ($(sf3.west)$) ;
\draw[-{Latex[length=1mm, width=0.8mm]}] ($(ffn2.east)$) -- ($(sf2.west)$) ;
\draw[-{Latex[length=1mm, width=0.8mm]}] ($(ffn1.east)$) -- ($(sf1.west)$) ;

\node[right of=sfn, node distance=1.4cm] (scn) {\LARGE $Score(i_n)$};
\node[right of=sf0, node distance=1.4cm] (sc0) { $\vdots$};
\node[right of=sf3, node distance=1.4cm] (sc3) {\LARGE $Score(i_3)$};
\node[right of=sf2, node distance=1.4cm] (sc2) {\LARGE $Score(i_2)$};
\node[right of=sf1, node distance=1.4cm] (sc1) {\LARGE $Score(i_1)$};

\draw[-{Latex[length=1mm, width=0.8mm]}] ($(sfn.east)$) -- ($(scn.west)$) ;
\draw[-{Latex[length=1mm, width=0.8mm]}] ($(sf3.east)$) -- ($(sc3.west)$) ;
\draw[-{Latex[length=1mm, width=0.8mm]}] ($(sf2.east)$) -- ($(sc2.west)$) ;
\draw[-{Latex[length=1mm, width=0.8mm]}] ($(sf1.east)$) -- ($(sc1.west)$) ;

\node[inner sep=0pt, right of=sc2, node distance=2.0cm] (rn)
    {\includegraphics[width=18px,height=18px]{fig/shoe.png}};
\node[right of=sc0, node distance=2.0cm] (r0) {$\vdots$};
\node[inner sep=0pt, right of=scn, node distance=2.0cm] (r3) {\includegraphics[width=18px,height=18px]{fig/women.png}};

\node[inner sep=0pt, right of=sc3, node distance=2.0cm] (r2) {\includegraphics[width=18px,height=18px]{fig/baby.png}};

\node[inner sep=0pt, right of=sc1, node distance=2.0cm] (r1) {\includegraphics[width=18px,height=18px]{fig/sports.png}};
\draw[-{Latex[length=1mm, width=0.8mm]}] ($(scn.east)$) -- ($(rn.west)$) ;
\draw[-{Latex[length=1mm, width=0.8mm]}] ($(sc3.east)$) -- ($(r3.west)$) ;
\draw[-{Latex[length=1mm, width=0.8mm]}] ($(sc2.east)$) -- ($(r2.west)$) ;
\draw[-{Latex[length=1mm, width=0.8mm]}] ($(sc1.east)$) -- ($(r1.west)$) ;

\node[above of=att1, node distance=0.8cm] (cp4) {\LARGE Attention};
\node[above of=sf1, node distance=0.8cm] (cp6) {\LARGE Softmax};

\node[] at ($(attn.east)+(0.6,-0.9)$) (cp7) {\LARGE $N_x$ blocks of Transformer encoder.};

 \draw[thick,dotted]  ($(en.south east)+(1.0, -0.15)$) rectangle ($(ffn1.north east)+(0.4, 0.1)$);

\draw [decorate,decoration={brace,amplitude=4pt},xshift=0pt,yshift=0pt]
($(in)+(0.4,-1.4)$) -- ($(in)+(-0.6,-1.4)$) node [black,midway,xshift=0cm,yshift=-0.6cm,align=center] 
{\LARGE Initial\\ \LARGE List};
\draw [decorate,decoration={brace,amplitude=4pt},xshift=0pt,yshift=0pt]
($(plusn)+(1.8,-1.4)$) -- ($(in)+(0.4,-1.4)$) node [black,midway,xshift=0cm,yshift=-0.6cm,align=center] 
{\LARGE Input\\ \LARGE Layer};
\draw [decorate,decoration={brace,amplitude=4pt},xshift=0pt,yshift=0pt]
($(ffn_n)+(0.8,-1.4)$) -- ($(plusn)+(1.8,-1.4)$) node [black,midway,xshift=0cm,yshift=-0.6cm,align=center] 
{\LARGE Encoding\\ \LARGE Layer};
\draw [decorate,decoration={brace,amplitude=4pt},xshift=0pt,yshift=0pt]
($(scn)+(0.55,-1.4)$) -- ($(ffn_n)+(0.8,-1.4)$) node [black,midway,xshift=0cm,yshift=-0.6cm,align=center] 
{\LARGE Output\\ \LARGE Layer};

\draw [decorate,decoration={brace,amplitude=4pt},xshift=0pt,yshift=0pt]
($(r3)+(0.45,-1.4)$) -- ($(scn)+(0.55,-1.4)$) node [black,midway,xshift=0cm,yshift=-0.6cm,align=center] 
{\LARGE Re-ranked\\ \LARGE List};


\node[dropout, left of=in, node distance=3.5cm, align=center] (do1) {Dropout};
\node[norm, above of=do1, node distance=0.7cm, align=center, font=] (n1) {Add \& Norm};
\node[mh, below of=do1, node distance=0.9cm, align=center] (trm1) {Multi-Head\\ Attention};

\draw[-{Latex[length=1mm, width=0.8mm]}] (trm1) edge (do1);
\draw[-{Latex[length=1mm, width=0.8mm]}] (do1) edge (n1);

\node[ff, above of=n1, node distance=1.6cm, align=center, font=] (ff1) {Feed Forward};
\node[dropout, above of=ff1, node distance=0.70cm, align=center, font=] (do2) {Dropout};
\node[norm, above of=do2, node distance=0.65cm, align=center, font=] (n2) {Add \& Norm};

\draw[-{Latex[length=1mm, width=0.8mm]}] (n1) edge (ff1);
\draw[-{Latex[length=1mm, width=0.8mm]}] (ff1) edge (do2);
\draw[-{Latex[length=1mm, width=0.8mm]}] (do2) edge (n2);

\node[below of=trm1, node distance=1.1cm, align=center] (input) {$\mathbf{E}$};

\node[above of=n2, node distance=1.0cm, align=center, font=\scriptsize] (out) {};

\draw[-{Latex[length=1mm, width=0.8mm]}] (n2) edge (out);
\draw[-{Latex[length=1mm, width=0.8mm]}] (input) edge (trm1);

\draw[-{Latex[length=1mm, width=0.8mm]}] ($(input.north)-(0,-0.1)$) -- ++(-1.2, 0) |- (n1.west) ;
\draw[-{Latex[length=1mm, width=0.8mm]}] ($(ff1.south)+(0,-0.2)$) -- ++(-1.2, 0) |- (n2.west) ;

\node[] (feedforward) at ($(n2)+(-1.3,0.6)$) {FFN};
\node[] (attention) at ($(n1)+(-1.0,0.6)$) {Attention};

\begin{pgfonlayer}{bg}
 \draw[thick,dotted]  ($(trm1.south west)+(-1.2, -1.0)$) rectangle ($(n2.north east)+(1.0, 0.8)$);
 \fill[thick,dotted,fill=yellow3,rounded corners=10pt]  ($(ff1.south west)+(-0.70, -0.4)$) rectangle ($(n2.north east)+(0.65, 0.2)$);
 \fill[thick,dotted,fill=blue0,rounded corners=10pt]  ($(trm1.south west)+(-0.85, -0.5)$) rectangle ($(n1.north east)+(0.65, 0.2)$);
\end{pgfonlayer};

\node[] at ($(attn.east)+(0.6,-0.6)$) (cp_huafan) {};
\node[pretrain_item1, right of=cp_huafan, node distance=8.5cm] (pretrain_e1) {};
\node[pretrain_item2, right of=pretrain_e1, node distance=1.5cm] (pretrain_e2) {};
\node[pretrain_item3, right of=pretrain_e2, node distance=1.5cm] (pretrain_e3) {};

\node[relu, above of=pretrain_e2, node distance=1.0cm] (pretrain_e4) {Layer$_1$};
\node[above of=pretrain_e4, node distance=0.7cm] (pretrain_e0) {$\vdots$};
\node[relu2, above of=pretrain_e0, node distance=0.5cm] (pretrain_e5) {Layer$_n$};
\node[sigmoid, above of=pretrain_e5, node distance=1.0cm] (pretrain_e6) {sigmoid};

\node[pvector, right of=pretrain_e6, node distance=2cm] (pv) {$\bm{pv_{i}}$};
\draw [->,thick,dotted] (pretrain_e5.east) to [in = -100, out = 0] (pv.south);

\node[below of=pretrain_e1, node distance=1cm] (input1) {\LARGE $\mathcal{H}_u$ };
\node[below of=pretrain_e2, node distance=1cm] (input2) {\LARGE item $i$};
\node[below of=pretrain_e3, node distance=1cm] (input3) {\LARGE user $u$};

\node[above of=pretrain_e6, node distance=1cm] (output) {\LARGE $P(y_{i}|\mathcal{H}_u, u;\theta^{'})$};

\draw[-{Latex[length=1mm, width=0.8mm]}] ($(input1.north)$) -- ($(pretrain_e1.south)$) ;
\draw[-{Latex[length=1mm, width=0.8mm]}] ($(input2.north)$) -- ($(pretrain_e2.south)$) ;
\draw[-{Latex[length=1mm, width=0.8mm]}] ($(input3.north)$) -- ($(pretrain_e3.south)$) ;

\draw[-{Latex[length=1mm, width=0.8mm]}] ($(pretrain_e2.north)$) -- ($(pretrain_e4.south)$) ;
\draw[-{Latex[length=1mm, width=0.8mm]}] ($(pretrain_e5.north)$) -- ($(pretrain_e6.south)$) ;
\draw[-{Latex[length=1mm, width=0.8mm]}] ($(pretrain_e6.north)$) -- ($(output.south)$) ;

 \draw[thick,dotted]  ($(pretrain_e1.south west)+(-0.5, -1.2)$) rectangle ($(pv.north east)+(0.3, 1.3)$);

\node[below of=input, node distance=1.6cm, align=center] (captiona) {\huge (a) One block of \\ \huge Transformer encoder.};
\node[right of=captiona, node distance=9.6cm, align=center] (captionb) {\huge (b) Architecture of PRM.};
\node[right of=captionb, node distance=11.6cm, align=center] (captionc) {\huge (c) The pre-trained model to \\ \huge generate $\bm{pv_{i}}, i=i_1,...,i_n$.};

\end{tikzpicture}
}
    \caption{The detailed network structure of our \textbf{PRM} (\textbf{P}ersonalized \textbf{R}e-ranking Model) and its sub-modules. }
    \label{fig:model1}
\end{figure*}

\subsection{Input Layer}
The goal of the input layer is to prepare comprehensive representations of all items in the initial list and feed it to the encoding layer. First we have a fixed length of initial sequential list $\mathcal{S} = [i_1, i_2, ... , i_n]$ given by the previous ranking method. Same as the previous ranking method, we have a raw feature matrix $\bm{X} \in \mathbb{R}^{n \times d_{\text{feature}}}$. Each row in $\bm{X}$ represents the raw feature vector $\bm{x_i}$ for each item $i \in \mathcal{S}$.

\textbf{Personalized Vector ($\bm{PV}$)}.
Encoding the feature vectors of two items can model the mutual influences between them, but to which extent these influences may affect the user is unknown. A user-specific encoding function need to be learned. Though the representation of the whole initial list can partly reflects the user's preferences, it is not enough for a powerful personalized encoding function. As shown in Figure~\ref{fig:model1} (b), we concat the raw feature matrix $\bm{X} \in \mathbb{R}^{n \times d_{\text{feature}}}$ with a personalized matrix $\bm{PV} \in \mathbb{R}^{n \times d_{\text{pv}}}$ to get the intermediate embedding matrix $\bm{E}^{'} \in \mathbb{R}^{n
 \times (d_{\text{feature}}+d_{pv})}$, which is shown in Equation~\ref{eq:concat_pv}. $\bm{PV}$ is produced by a pre-trained model which will be introduced in the following section. The performance gain of $\bm{PV}$ will be introduced in the evaluation section. 
\begin{equation}
	\bm{E}^{'} = 
\begin{bmatrix}
\bm{x_{i_1}} \, ; \, \bm{pv_{i_1}}\\
\bm{x_{i_2}} \, ; \, \bm{pv_{i_2}}\\
\dots  \\
\bm{x_{i_n}} \, ; \, \bm{pv_{i_n}}\\
\end{bmatrix}
	\label{eq:concat_pv}
\end{equation}

\textbf{Position Embedding ($\bm{PE}$)}. In order to utilize the sequential information in the initial list, we inject a position embedding $\bm{PE} \in \mathbb{R}^{n
 \times (d_{\text{feature}}+d_{pv})}$ into the input embedding. Then the embedding matrix for encoding layer can be calculated using Equation~\ref{eq:pe}. In this paper, a learnable $\bm{PE}$ is used which we  found that it slightly outperforms the fixed position embedding used in \cite{vaswani2017attention}.
\begin{equation}
	\bm{E}^{''} = 
\begin{bmatrix}
\bm{x_{i_1}} \, ; \, \bm{pv_{i_1}}\\
\bm{x_{i_2}} \, ; \, \bm{pv_{i_2}}\\
\cdots \\
\bm{x_{i_n}} \, ; \, \bm{pv_{i_n}}\\
\end{bmatrix}
+
\begin{bmatrix}
\bm{pe_{i_1}} \\
\bm{pe_{i_2}} \\
\cdots\\
\bm{pe_{i_n}} \\
\end{bmatrix}
	\label{eq:pe}
\end{equation}


At last we use one simple feed-forward network to convert the feature matrix $\bm{E}^{''} \in \mathbb{R}^{n
 \times (d_{\text{feature}}+d_{pv})}$ to $\bm{E} \in \mathbb{R}^{n \times d}$, where $d$ is latent dimensionality of each input vector of encoding layer. $\bm{E}$ can be formulated as Equation~\ref{eq:linear}.
\begin{equation}
	\bm{E} = \bm{E}\bm{W}^E + b^E
	\label{eq:linear}
\end{equation}
where $\bm{W}^E \in \mathbb{R}^{(d_{\text{feature}}+d_{pv}) \times d}$ is the projection matrix and $b^E$ is $d$-dimensional vector.

\subsection{Encoding Layer}
The goal of the encoding layer in Figure~\ref{fig:model1}(a) is to integrate the mutual influences of item-pairs and other extra information, includes the user preferences and the ranking order of the initial list $\mathcal{S}$. To achieve this goal, we adopt Transformer-like encoder because Transformer\cite{vaswani2017attention} has been proven to be effective in many NLP tasks, specially in machine translation for its powerful encoding and decoding ability compared to RNN-based approaches\cite{cho2014learning,Hochreiter:LSTM:NC1997,cho2014properties}. The self-attention mechanism in Transformer is particularly suitable in our re-ranking task as it directly models the mutual influences for any two items regardless the distances between them. Without distance decay, Transformer can capture more interactions between items that are far away from each other in the initial list. As shown in Figure~\ref{fig:model1}(b), our encoding module consists of $N_x$ blocks of Transformer encoder. Each block (Figure~\ref{fig:model1}(a)) contains an attention layer and a Feed-Forward Network (FFN) layer. 

\textbf{Attention Layer}.
The attention function we used in this paper is defined as Equation~\ref{eq:sf}:
\begin{equation}
\text{Attention}(\bm{Q,K,V}) = \text{softmax}\left(\frac{\bm{QK}^T}{\sqrt{d}}\right)\bm{V},
	\label{eq:sf}
\end{equation}
where matrices $\bm{Q,K,V}$ represent queries, keys and values respectively. $d$ is the dimensionality of matrix $\bm{K}$ to avoid large value of the inner product. \textit{softmax} is used to convert the value of inner-product into the adding weight of the value vector $\bm{V}$. In our paper, we use self-attention where $\bm{Q,K}$ and $\bm{V}$ are projected from the same matrices. 

To model more complex mutual influences, we use the multi-head attention as shown in Equation~\ref{eq:mh}: 
\begin{equation}
\begin{split}
\bm{S}^{'}  = \text{MH}(\bm{E}) & = \text{Concat}(\text{head}_1,...,\text{head}_h)\bm{W^O} \\
  \text{head}_i &= \text{Attention}(\bm{E}\bm{W}^Q, \bm{E}\bm{W}^K, \bm{E}\bm{W}^V),
\end{split}
	\label{eq:mh}
\end{equation}
where $\bm{W}^Q,\bm{W}^K,\bm{W}^V \in \mathbb{R}^{d \times d}$. $\bm{W^O} \in \mathbb{R}^{hd \times d_{\text{model}}}$ is the projection matrix. $h$ is the number of headers. The influence of different value of $h$ will be studied in the ablation study in the next section.  

\textbf{Feed-Forward Network}.
The function of this position-wise Feed-Forward Network (FFN) is mainly to enhance the model with non-linearity and interactions between different dimensions of the input vectors. 

\textbf{Stacking the Encoding Layer}. Here we use attention module followed by the position-wise FFN as a block of Transformer\cite{vaswani2017attention} encoder. By stacking multiple blocks, we can get more complex and high-order mutual information. 

\subsection{Output Layer}

The function of the output layer is mainly to generate a score for each item $i=i_1,\dots,i_n$ (labeled as $Score(i)$ in Figure~\ref{fig:model1} (b)). We use one linear layer followed by a softmax layer. The output of softmax layer is the probability of click for each item, which is labeled as $P(y_{i} | \bm{X}, \bm{PV};\hat{\theta})$. We use $P(y_{i} | \bm{X}, \bm{PV};\hat{\theta})$ as $Score(i)$ to re-rank the items in one-step. The formulation of $Score(i)$ is: 
\begin{equation}
\begin{split}
	Score(i)  = P(y_{i} | \bm{X}, \bm{PV}; \hat{\theta})
	 = \text{softmax} \Bigl( \bm{F^{(N_x)}} \bm{W}^{F} + \bm{b}^{F} \Bigr), i \in \mathcal{S}_r
\end{split}
	\label{eq:prediction}
\end{equation}
where $\bm{F^{(N_x)}}$ is the output of $N_x$ blocks of Transformer encoder. $\bm{W}^{F}$ is learnable projection matrix, and $\bm{b}^{F}$ is the bias term. $n$ is the number of items in the initial list. 

In the training process, we use the click-through data as label and minimize the loss function shown in Equation~\ref{eq:loss}. 

\begin{equation}
\begin{aligned}
	\mathcal{L} = -\sum_{r \in \mathcal{R}} \sum_{i \in \mathcal{S}_r} & {y_{i} \log(P(y_{i}|\bm{X}, \bm{PV};\hat{\theta}) } \\
\end{aligned}
	\label{eq:loss}
\end{equation}

\subsection{Personalized Module}
In this section, we introduce the approach to calculate the personalized matrix $\bm{PV}$, which represents interactions between user and items. The straightforward approach is to learn $\bm{PV}$ with PRM model in an end-to-end manner via the re-ranking loss. However, as explained in Section\ref{sec:formulation}, the re-ranking task is to refine the output of previous ranking approaches. The task-specific representation learned on re-ranking task lacks users' generic preferences. Therefore, we utilize a pre-trained neural network to produce user's personalized embeddings $\bm{PV}$ which are then used as extra features for PRM model. The pre-trained neural network is learned from the whole click-through logs of the platform. 
Figure~\ref{fig:model1}(c) shows the structure of pre-trained model used in our paper. This sigmoid layer outputs the click probability ($P(y_{i}|\mathcal{H}_u, u;\theta^{'})$)  on item $i$ for user $u$ given user's all behavior history ($\mathcal{H}_u$) and the side information of the user. 
The side information of user includes  \textit{gender, age} and \textit{purchasing level}, et.al. The loss of the model is calculated by a point-wise cross entropy function which is shown in Equation~\ref{eq:pretrain-loss}.
\begin{equation}
\begin{aligned}
	\mathcal{L} = \sum_{i \in \mathcal{D}} & {(y_{i} \log(P(y_{i}|\mathcal{H}_u, u;\theta^{'})) } \\
	& + (1-y_{i}) \log(1- P(y_{i}|\mathcal{H}_u, u;\theta^{'})),
\end{aligned}
	\label{eq:pretrain-loss}
\end{equation}
where $\mathcal{D}$ is the set of items displayed to user $u$ on the platform. $\theta^{'}$ is the parameter matrix of pre-trained model. 
$y_i$ is the label (click or not) on item $i$.
Inspired by the work\cite{covington2016deep}, we employ the hidden vector before the \textit{sigmoid} layer as the personalized vector $\bm{pv_i}$ (in Figure~\ref{fig:model1}(c)) that feeds into our PRM model.

Figure~\ref{fig:model1}(c) shows one possible architecture of the pre-trained model, other general models such as FM\cite{Rendle:2010:FM:1933307.1934620}, FFM\cite{Liu:2018:FPE:3240323.3240396}, DeepFM\cite{Guo:2017:DFB:3172077.3172127}, DCN\cite{Wang:2017:DCN:3124749.3124754}, FNN\cite{zhou2016deep} and PNN\cite{Qu:2018:PNN:3289475.3233770} can also be used as alternatives to generate $\bm{PV}$.

 
\section{Experimental Resutls}
In this section, we first introduce the datasets and baselines used for evaluation. Then we compare our methods with baselines on these datasets to evaluate the effectiveness of our PRM model. At the same time, the ablation study is conducted to help understand which part of our model contributes most to the overall performance.
 
\subsection{Datasets}
We evaluate our approach based on two datasets: Yahoo! Webscope v2.0 set 1\footnote{http://webscope.sandbox.yahoo.com} (abbreviated as Yahoo Letor dataset) and E-commerce Re-ranking dataset\footnote{Our dataset is available at \url{https://github.com/rank2rec/rerank}.} To the best of our knowledge, there is no publically available re-ranking dataset with context information for recommendation. Therefore, we construct E-commerce Re-ranking dataset from a popular e-commerce platform. The overview of two datasets are shown in Table~\ref{tab:dataset}.


\textbf{Yahoo Letor dataset}. We process the Yahoo Letor dataset to be fit for the ranking model of recommendation using the same method in Seq2Slate\cite{bello2018seq2slate}.  Firstly, we convert the ratings ($0$ to $4$) to binary labels using a threshold $T_b$. Secondly, we use a decay factor $\eta$ to simulate the impression probabilities of items. All the documents in Yahoo Letor dataset are rated by the experts under the assumption that all documents for each query can be viewed by the users compeletely. However, in the real world recommendation scenario, items are viewed by the users in a top-down manner. As the screen of the mobile App can only show limited number of items, the higher the ranked position of one item, the smaller probability of that this item can be viewed by the user. In this paper, we use $1/pos(i)^\eta$ as the decay probability, where $pos(i)$ is the ranking position of item $i$ in the initial list.

\textbf{E-commerce Re-ranking dataset}. The dataset contains a large-scale records in form of click-through data from a real world recommendation system. Each record in the dataset contains a recommendation list for each user with users' basic information, click-through labels and raw features for ranking. 


\begin{table}[t]
\caption{Overview of the datasets.}
\centering
\begin{tabular}{lp{1.8cm}p{2.5cm}}
\toprule
  & Yahoo$\,$Letor Dataset & E-commerce Re-ranking Dataset \\
\midrule
$\#$Users & - & 743,720  \\
$\#$Docs/Items & 709,877 & 7,246,323  \\
$\#$Records & 29,921 & 14,350,968\\
Relavance/Feedback & \{0,1,2,3,4\} & \{0,1\} \\
\bottomrule
\end{tabular}
\label{tab:dataset}
\end{table}
     
\subsection{Baselines}
Both learning to rank (LTR) and re-ranking methods can act as our baselines. 

\textbf{LTR}. The LTR methods are used in two tasks. Firstly, the LTR methods can generate an initial list $\mathcal{S}_r$ for the re-ranking model from a candidate set $\mathcal{I}_r$ for each user request $r$. Secondly, the LTR methods which use pairwise or listwise loss function can act as re-ranking methods by taking the initial list $\mathcal{S}_r$ as input and conducting the ranking algorithm for another time. The representative LTR methods used in this paper include:
\begin{itemize}
	\item SVMRank\cite{joachims2006training}: This is a representative learning to rank method which use the pairwise loss to model the scoring function. 
	\item LambdaMart\cite{burges2010ranknet}: This is a representative learning to rank method which use the listwise loss to model the scoring function. LambdaMart is the state-of-the-art LTR among those LTR methods equipped with the listwise loss function according to \cite{Wu:2018:TCP:3209978.3209993}'s evaluation. 
	\item DNN-based LTR: This is the learning to rank method which is deployed in our online recommender system. It use the standard Wide\&Deep network structure\cite{Cheng:2016:WDL:2988450.2988454} to model the scoring function via the pointwise loss function. 
\end{itemize}

\textbf{Re-ranking}. As mentioned in the related work section, the existing re-ranking methods include DLCM\cite{ai2018learning}, Seq2Slate\cite{bello2018seq2slate} and GlobalRerank\cite{zhuang2018globally}. DLCM\cite{ai2018learning} and GlobalRerank\cite{zhuang2018globally} focus on re-ranking in information retrieval. Seq2Slate\cite{bello2018seq2slate} focuses on re-ranking in both recommendation and information retrieval. In this paper, we only choose DLCM as baseline method. Seq2Slate and GlobalRerank are not chosen as baselines because they all use the decoder structure to generate the re-ranked list. Seq2Slate uses pointer network to generate re-ranked list sequentially. GlobalRerank uses RNN equipped with attention mechanism as the decoder. The decoder structure outputs the item one by one. Whether an item is selected depends on the items which are chosen before it. As a consequence, both Seq2Slate and GlobalRerank can not be parallelized in online inference. The time complexity for Seq2Slate and GlobalRerank at interference phase is $\mathbf{O}(n)\times RT$, where $n$ is the length of the initial list and $RT$ is the time for a single ranking or re-ranking request. The latency for re-ranking by Seq2Slate and GlobalRerank is unacceptable because of the strict latency criterion for online recommender service. 
\begin{itemize}
	\item DLCM\cite{ai2018learning}: It is a re-ranking model used in information retrieval based on the initial list generated by LTR methods. The GRU is used to encode the local context information into a global vector. Combing the global vector and each feature vector, it learns a more powerful scoring function than the global ranking function of LTR.
\end{itemize}


\subsection{Evaluation Metrics}
For offline evaluation, we use \textit{Precision} and \textit{MAP} to compare different methods. More specifically, we use Precision$@$5, Precision$@$10 for precision and MAP$@$5, MAP$@$10 and MAP$@$30 for MAP. As the maximum length of initial list in our experiments is 30, MAP$@$30 represents total MAP and is denoted by MAP in this paper. The definitions of the metrics are as follows.

\textbf{Precision$@$k} is defined as the the fraction of clicked items in the top-k recommended items for all test samples, as shown in Equation~\ref{eq:precision_k}.

\begin{equation}
	Precision@k = \frac{1}{|\mathcal{R}|} \sum_{r \in \mathcal{R}} {\frac{\sum_{i=1}^{k}{\mathbbm{1}(\mathcal{S}_r(i))}}{k}} 
	\label{eq:precision_k}
\end{equation} 

 where $\mathcal{R}$ is the set of all user requests in the test dataset. $\mathcal{S}_r$ is the ordered list of items given by the re-ranking model for each request $r \in \mathcal{R}$ and $\mathcal{S}_r(i)$ is the $i$-th item. $\mathbbm{1}$ is the indicator function whether item $i$ is clicked or not. 

\textbf{MAP$@$k} is short for the mean average precision of all ranked lists cut off by $k$ in the test dataset. It is defined as follows.
\begin{equation}
	MAP@k = \frac{1}{|\mathcal{R}|} \sum_{r \in \mathcal{R}}{\frac{\sum_{i=1}^k{Precision@i*\mathbbm{1}(\mathcal{S}_r(i))}}{k}} 
	\label{eq:map_k}
\end{equation}    

For online A/B test, we use PV, IPV, CTR and GMV as metrics. PV and IPV are defined as the total number of items viewed and clicked by the users. CTR is the clickthrough rate and can be calculated by IPV/PV. GMV is the total amount of money (revenue) user spent on the recommended items.

\subsection{Experimental Settings}

For both baselines and our PRM model, we use the same value for those critical hyper parameters. The hidden dimensionality $d_{\text{model}}$ is set to $1024$ for Yahoo Letor dataset and $64$ for E-commerce Re-ranking dataset. The learning rate of Adam optimizer in our PRM model is the same with \cite{vaswani2017attention}. Negative log likelihood loss function is used as shown in Equation~\ref{eq:loss}. $p_{\text{dropout}}$ is set to $0.1$. The batch size is set to 256 for Yahoo Letor dataset and 512 for E-commerce Re-ranking dataset. These settings are got by fine-tuning the baselines to achieve better performance. We also try different experimental settings, the results are consistent with the current settings and are ommited. The rest of the settings belonging to the customized parts of our model will be listed at the corresponding parts in the evaluation section.  

\subsection{Offline Experiments}
In this section, we first conduct offline evaluations on Yahoo Letor dataset and E-commerce Re-ranking dataset. Then we show the results of online A/B test. We also conduct the ablation study to help finding which part of our PRM model contributes most to the performance.

\begin{table*}[h]
\caption{Offline evaluation results on Yahoo Letor dataset.}
\centering
\begin{tabular}{lcccccc}
\hline
\multirow{2}{*}{Init. List} & \multirow{2}{*}{Reranking}  & \multicolumn{5}{c}{Yahoo Letor dataset.} \\
\cline{3-7}
  & &  Precision$@$5(\%) & Precision$@$10(\%) &  MAP$@$5(\%) & MAP$@$10(\%) & MAP(\%) \\ 
\hline
\multirow{4}{*}{SVMRank} & SVMRank & 50.42 & 42.25 & 73.71 & 68.28 & 62.14 \\
& LambdaMART & 51.35 & 43.08 & 74.94 & 69.54 & 63.38 \\
& DLCM   & 52.54  &  43.26 & 76.52 & 70.86 & 64.50  \\
 & PRM-BASE  &  \textbf{53.29}  &  \textbf{43.66} & \textbf{77.62} & \textbf{72.02} & \textbf{65.60}  \\
\hline
 \multirow{4}{*}{LambdaMART} & SVMRank   & 50.41 & 42.34 & 73.82 & 68.27 & 62.13 \\
 & LambdaMART & 52.04 & 43.00 & 75.77 & 70.49 & 64.04 \\
 & DLCM     & 52.54  & 43.16  & 77.81 & 71.88 & 65.24 \\
 & PRM-BASE   & \textbf{53.63}  & \textbf{43.41}  & \textbf{78.62} & \textbf{72.67} & \textbf{65.72}  \\
 \hline
\end{tabular}
\label{tab:offline-yahoo}
\end{table*}

\begin{table}[h]
\caption{Ablation study of PRM-BASE on Yahoo Letor datasets with the initial list generated by SVMRank. All the numbers in the table are multiplied by 100.}
\centering
\begin{tabular}{lrrrrr}
\toprule
  & \multicolumn{4}{c}{Yahoo Letor dataset}  \\
 \cmidrule{2-6}
 & P$@$5 & P$@$10 & MAP$@$5 & MAP$@$10 & MAP \\ 
\midrule
DLCM & 52.54  &  43.26 & 76.52 & 70.86 & 64.50  \\
\midrule
Default(b=4,h=3) & 53.29 & 43.66 & 77.62 & 72.02 & 65.60 \\
\midrule
Remove PE & \textbf{52.55} & \textbf{43.56} & 76.11 & 70.74 & \textbf{64.73} \\
Remove RC & 53.24 & 43.63 & 77.52 & 71.92 & 65.52 \\
Remove Dropout & 53.17 & 43.42 & 77.41 & 71.80 & 65.17 \\
Block(b=1) & 53.12 & 43.59 & 77.58 & 71.91 & 65.49 \\
Block(b=2) & 53.19 & 43.58 & 77.51 & 71.86 & 65.49 \\
Block(b=6) & 53.22 & 43.63 & 77.64 & 72.02 & 65.61\\
Block(b=8) & \textbf{52.85} & \textbf{43.32} & \textbf{77.43} & \textbf{71.65} & \textbf{65.14}\\
Multiheads(h=1) &  53.17 & 43.67 & 77.65 & 71.96 & 65.55\\
Multiheads(h=2) & 53.29 & 43.60 & 77.68 & 72.00 & 65.57 \\
Multiheads(h=4) & 53.20 & 43.61 & 77.72 & 72.00 & 65.58 \\
\bottomrule
\end{tabular}
\label{tab:ablation-yahoo}
\end{table}

\begin{table*}[t]
\caption{Offline evaluation results on E-commerce Re-ranking dataset.}
\centering
\begin{tabular}{lcrrrrr}
\hline
\multirow{2}{*}{Init. List} & \multirow{2}{*}{Re-ranking}  & \multicolumn{5}{c}{E-commerce Re-ranking dataset.} \\
\cline{3-7}
  & & Precision$@$5 & Precision$@$10 &  MAP$@$5(\%) & MAP$@$10(\%) & MAP(\%) \\ 
  \hline
\multirow{3}{*}{DNN-based LTR} & DLCM    &  12.21  & 9.73 & 29.32 & 30.28 & 28.19 \\
 & PRM-BASE    &  12.71 & 9.99 & 29.80 & 30.83 & 28.85 \\
  & PRM-Personalized-Pretrain  &  \textbf{13.58} & \textbf{10.52} & \textbf{31.18} & \textbf{32.12} & \textbf{30.15} \\
\hline
\end{tabular}
\label{tab:offline-our}
\end{table*}

\begin{table}[h]
\vspace{7pt}
\caption{Performance improvements in online A/B test compared with a DNN-based LTR without re-ranking method.}
\centering
\begin{adjustbox}{width=\linewidth}
\begin{tabular}{lrrrr}
\toprule
 Reranking & PV & IPV & CTR & GMV \\ 
\midrule
 DLCM        & 0.77$\%$  & 1.75$\%$  & 0.97$\%$ & 0.13$\%$  \\
 PRM-BASE    & 1.27$\%$  & 2.44$\%$  & 1.16$\%$ & 0.36$\%$  \\
 PRM-Personalized-Pretrain  & \textbf{3.01$\%$} & \textbf{5.69$\%$} &\textbf{2.6$\%$} & \textbf{6.65$\%$} \\
\bottomrule
\end{tabular}
\end{adjustbox}
\label{tab:online}
\end{table}

\subsubsection{Offline Evaluation on Yahoo Letor dataset.}
In this section, we conduct evaluation on Yahoo Letor dataset to discuss the following questions:

\begin{itemize}
	\item RQ0: Does our PRM model outperform the state-of-the-art methods and why?
	\item RQ1: Does the performance vary according to initial lists generated by different LTR approaches?  
\end{itemize}

 The evaluation results are shown in Table~\ref{tab:offline-yahoo}. We compare the baselines and our PRM-BASE model based on two different initial lists which are generated by LambdaMART and SVMRank respectively. PRM-BASE is the variant of our PRM model without the personalized module.  Note that Yahoo Letor dataset does not contain user-related information, thus we only conduct PRM-BASE for comparison. SVMRank and LambdaMart are also used for re-ranking. For SVMRank, we use the implementation in \cite{joachims2006training}. For LambdaMart, we use the implementation from RankLib\footnote{\url{https://sourceforge.net/p/lemur/wiki/RankLib/}}. 

Table~\ref{tab:offline-yahoo} shows that our PRM-BASE  achieves stable and significant performance improvements comparing with all baselines. When based on the initial list generated by SVMRank, PRM-BASE outperforms DLCM by 1.7$\%$ at MAP and 1.4$\%$ at Precision$@$5. The gap gets larger when comparing with SVMRank which has 5.6$\%$ increase at MAP and 5.7$\%$ increase at Precision$@$5. When based on the initial list generated by LambdaMART, PRM-BASE outperforms DLCM by 0.7$\%$ at MAP and 2.1$\%$ at Precision$@$5. PRM-BASE also achieves 2.6$\%$ improvements on MAP and 3.1$\%$ improvements on Precision$@$5 comparing with LambdaMART . 

PRM-BASE uses the same training data as DLCM and does not contain the personalized module. The performance gain over DLCM mainly comes from the powerful encoding ability of Transformer. Multi-head attention mechanism performs better at modeling mutual influence between two items, especially when the length of encoding list gets longer\cite{kang2018self}. In our model, the attention mechanism can model the interactions of any item-pairs in $\mathbf{O}$(1) encoding distance. 

As PRM-BASE uses the Transformer-like structure, there are many sub-modules which may contribute to the performance. We conduct the ablation study to help us understand which sub-design helps the most to beat the baselines. The ablation study is conducted on the initial list generated by SVMRank. Similar results were found when using the initial list generated by LambdaMART and we omits the results in this paper as space is limited. Table~\ref{tab:ablation-yahoo} show the results of ablation in three parts: The first part (first row) shows the performance of the baseline DLCM. The second part (second row) ``Default'' is the best performance of our PRM-BASE model. The third part (the remaining rows) shows different ablation variants of our PRM model which include: remove position embedding (PE), remove residual connection (RC), remove dropout layer, use different number of blocks and use different number of heads in multi-head attention. Note that we set b=4 and h=3 in our ``Default'' PRM model. 

As shown in Table~\ref{tab:ablation-yahoo}, the performance of our model degrades greatly after removing position embedding. This confirms the importance of sequential information given by the initial list. After removing the position embedding, our model learns the scoring function from the candidate set instead of an ordered list. Note that even without position embedding, our PRM-BASE still achieve comparable performance with DLCM, which further confirms that our PRM-BASE model can encode the initial list more effectively than DLCM. 

The MAP of our model slightly decreases by 0.1$\%$ and 0.7$\%$ respectively when removing residual connections and dropout layer, which indicates that our model is less severe to the problems such as gradients vanishing and overfitting. The performance of our model first increases with the number of blocks (1->2->4) and decreases afterwards (4->6->8), as overfitting happens when we stack 8 encoding blocks together.

We also tried different settings($h=1,2,3,4$) in the multi-head attention layer. No significant improvements are observed in Table~\ref{tab:ablation-yahoo}, which is different from the conclusions derived from NLP tasks\cite{tang2018self}. The experiments in NLP show that when using more heads in multi-head attention mechanism, it is usually helpful since more information can be captured for the following reasons. (1) From Equation~\ref{eq:mh} we find that the function of each \textit{head} is playing a role of mapping the original feature vector into a different subspace. Thus using more heads, we can model more interactions of items in different sub-spaces. (2) \cite{tang2018self} indicates that using more heads is helpful in encoding the information of long sequence. This is reasonable because the output vector for a certain item is the weighted sum of all item vectors in the list. When the sequence becomes longer, each item in the list contributes less to the output vector. However, in our re-ranking settings, all the items in the initial list are highly homogenous. There are minor improvements when mapping the original feature vector into more different subspaces. As a consequence, we suggest to use only one head to save computation costs because the performance improvements are not obvious.





\subsubsection{Offline Evaluation on E-commerce Re-ranking dataset}

We conduct the offline evaluation on E-commerce Re-ranking dataset to answer the following question.

\begin{itemize}
	\item RQ2: What is the performance of our PRM model equipped with personalized module?
\end{itemize} 

The evaluation results are shown in Table~\ref{tab:offline-our}. For our PRM models, we not only evaluated the performance of PRM-BASE, but also evaluated the performance of the variant of model equipped with the pre-trained personalized vector $\bm{PV}$, which is labelled as PRM-Personalized-Pretrain. 
As our previous evaluation on Yahoo Letor dataset already confirms that our model and DLCM achieve better performance in all metrics and DLCM\cite{ai2018learning} also has consistent results, we omit the comparison with SVMRank and LambdaMART on our E-commerce Re-ranking dataset. The initial list is generated by a DNN-based LTR method which is deployed in our real world recommender system. 

Table~\ref{tab:offline-our} shows consistent results with Table~\ref{tab:offline-yahoo} when comparing PRM-BASE with DLCM. Our PRM-BASE outperforms DLCM by 2.3$\%$ at MAP and 4.1$\%$ at Precision$@$5. Recall that on Yahoo Letor dataset, PRM-BASE achieves 1.7$\%$ improvements on MAP and 1.4$\%$ improvements on Precision$@$5. The performance gain on our E-commerce Re-ranking dataset is much larger than on Yahoo Letor dataset. This is highly related with the properties of Yahoo Letor dataset. 
Our statistics of the Yahoo Letor dataset show that the average click through rate is 30$\%$, which mean that for each query with 30 recommended documents, about 9 documents are clicked by the users. However, the average click-through rate in our real world E-commerce Re-ranking dataset is no more than 5$\%$. It means that ranking on Yahoo Letor dataset is much easier than on E-commerce Re-ranking dataset. This is also confirmed by the value of MAP for the same ranking methods on two datasets: DLCM can achieve 0.64 MAP on Yahoo Letor dataset but can only achieve 0.28 MAP on E-commerce Re-ranking dataset. Combining Table~\ref{tab:offline-our} and Table~\ref{tab:offline-yahoo}, we find that the harder the ranking task, the larger improvements of our PRM model.

Table~\ref{tab:offline-our} shows that our PRM-Personalized-Pretrain achieves significant performance improvements comparing with PRM-BASE. PRM-Personalized-Pretrain outperforms PRM-BASE by 4.5$\%$ at MAP and 6.8$\%$ at Precision$@$5. This is mainly imported by the personalized vector $\bm{PV}$, which is learned by a pre-trained model whose architecture is illustrated in Figure~\ref{fig:model1} (c). PRM-Personalized-Pretrain has two advantages: (1) The pre-trained model can fully utilize longer period of users' logs to provide more generic and representative embeddings of users' preferences. (2) Equipped with long term and generic user embedding, our PRM model is able to learn better user-specific encoding function which can  more precisely capture mutual influences of item-pairs for each user. Note that the architecture of the pre-trained model is not highly coupled with our PRM model, other general models\cite{Wang:2017:DCN:3124749.3124754,Liu:2018:FPE:3240323.3240396,Guo:2017:DFB:3172077.3172127,zhou2016deep,Qu:2018:PNN:3289475.3233770} can also be used as alternatives to generate $\bm{PV}$.  

\subsection{Online Experiments}
We also conduct online A/B test at a real world e-commerce recommender system on online metrics which includes PV, IPV, CTR and GMV. The meaning of these metrics is explained in the previous ``Evaluation Metrics'' section. These metrics evaluate how much willingness for users to view (PV), click (IPV, CTR) and purchase (GMV) in a recommender system. For each algorithm, there are hundreds of thousands of users and millions of requests for online test. 

Table~\ref{tab:online} shows the relative improvements of three methods to an online base ranker (DNN-based LTR). Firstly, the online A/B test shows that re-ranking helps increase the online metrics no matter what kinds of re-ranking methods are. Again, we can conclude that re-ranking helps improving the performance by considering the mutual influences of items in the initial list. It is noteworthy that 0.77$\%$ increase (DLCM v.s. Without re-ranking) on PV is significant in our online system because it means that about billions of extra items are viewed by the users after using the re-ranking method. Secondly, we can conclude that our PRM-BASE model brings an extra 0.50$\%$ absolute increase on viewed items and extra 0.69$\%$ absolute increase on clicked items compared with DLCM. Lastly, by using the personalized module, our PRM-Personalized-Pretrain model can further improve the GMV by 6.29$\%$ absolute increase compared with PRM-BASE. Recall that in offline experiments on E-commerce Re-ranking dataset, PRM-Personalized-Pretrain has 4.5$\%$ increase at MAP compared with PRM-BASE. The result shows that personalized encoding function with pre-trained users' representations can help capture more precise interactions of item-pairs and bring significant performance gain for re-ranking method.

\subsection{Visualizing Attention Weights}
We visualize the attention weights learned by our model to answer the following question.

\begin{itemize}
	\item RQ3: Can self-attention mechanism learn meaningful information with respect to different aspects, for example, positions and characteristics of items?
\end{itemize} 

\begin{figure}
     \centering
     \subfloat[][Category.]{\includegraphics[width=.5\linewidth]{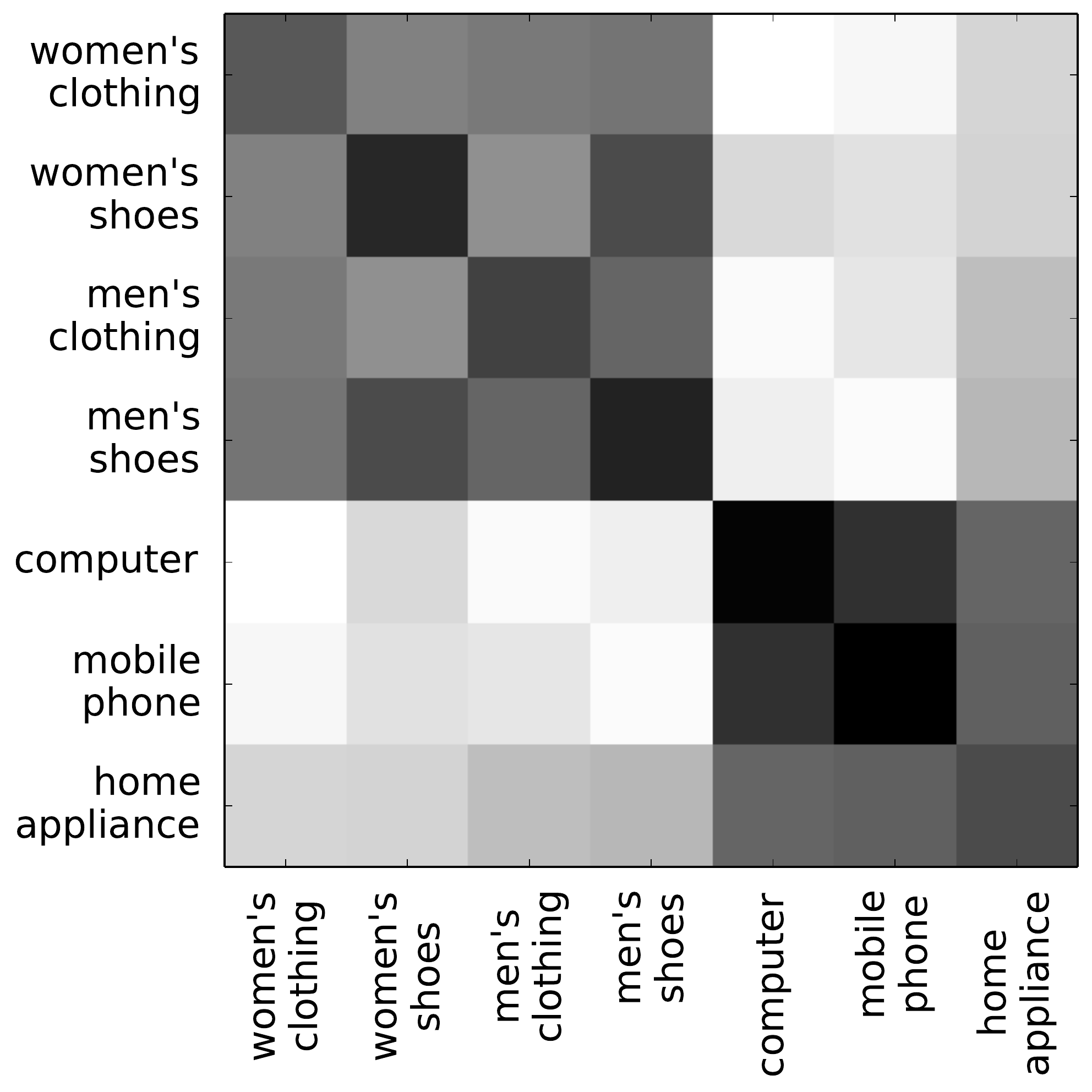}\label{<figure1>}}
     \subfloat[][Price.]{\includegraphics[width=.5\linewidth]{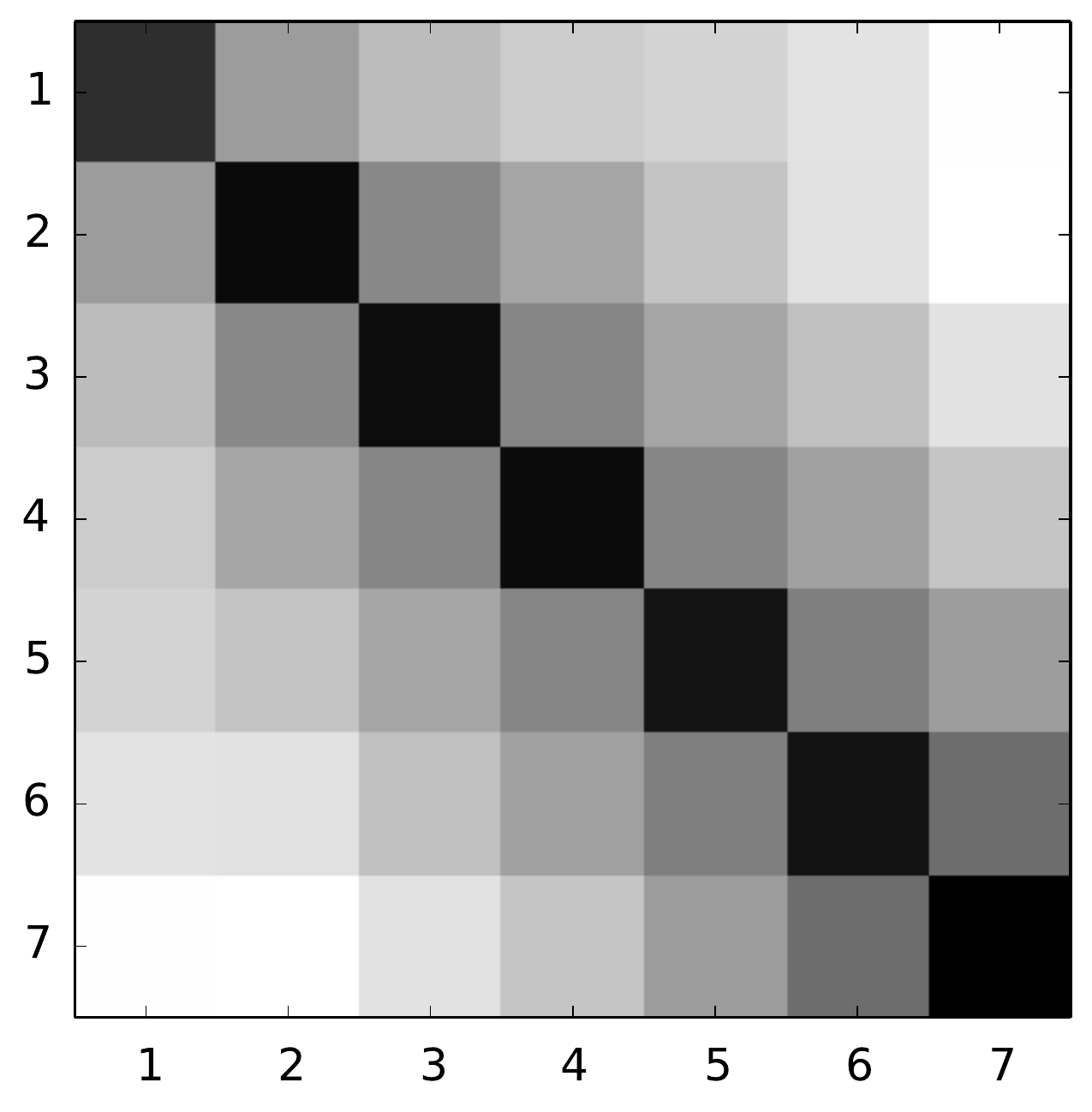}\label{<figure2>}}
     \caption{Average attention weights related to items' attributes.}
     \label{fig:items}
\end{figure}

\textbf{Attention on Characteristics.}
We first visualize the average attention weights between items on two characteristics: \textit{category} and \textit{price}. The results calculated on the test dataset are shown in Figure~\ref{fig:items}. Each block in the heatmap represents the average attention weights between items belonging to seven main categories. The darker the block, the larger the weight. From Figure~\ref{fig:items}(a) we can conclude that the attention mechanism can successfully capture mutual-influences in different categories. The items with similar categories tend to have larger attention weights, indicating larger mutual influences. For example, ``men's shoes'' has more influences on ``women's shoes'' than on ``computer''. It is also easy to understand that ``computer'', ``mobile phone'' and ``home appliance'' have large attention weights with each other because they are all electronics. Similar cases can be observed in Figure~\ref{fig:items}(b). In Figure~\ref{fig:items}(b), we classify the items into 7 levels according to their prices. The closer price between items, the larger the mutual influences.
     
\begin{figure}
     \centering
	 \subfloat[][With position embedding.]{\includegraphics[width=.5\linewidth]{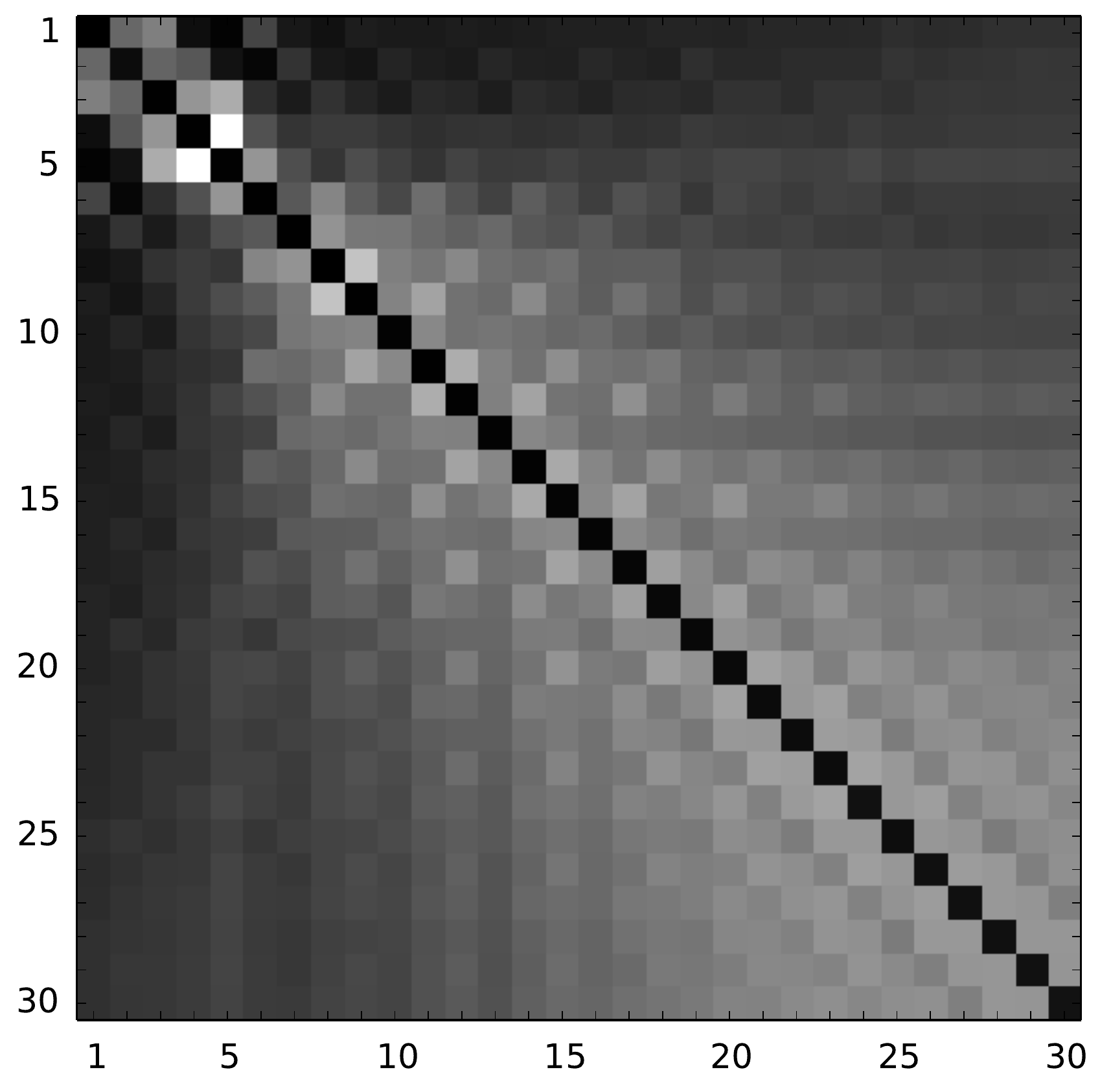}\label{<figure2>}}
	 \subfloat[][Without position embedding.]{\includegraphics[width=.5\linewidth]{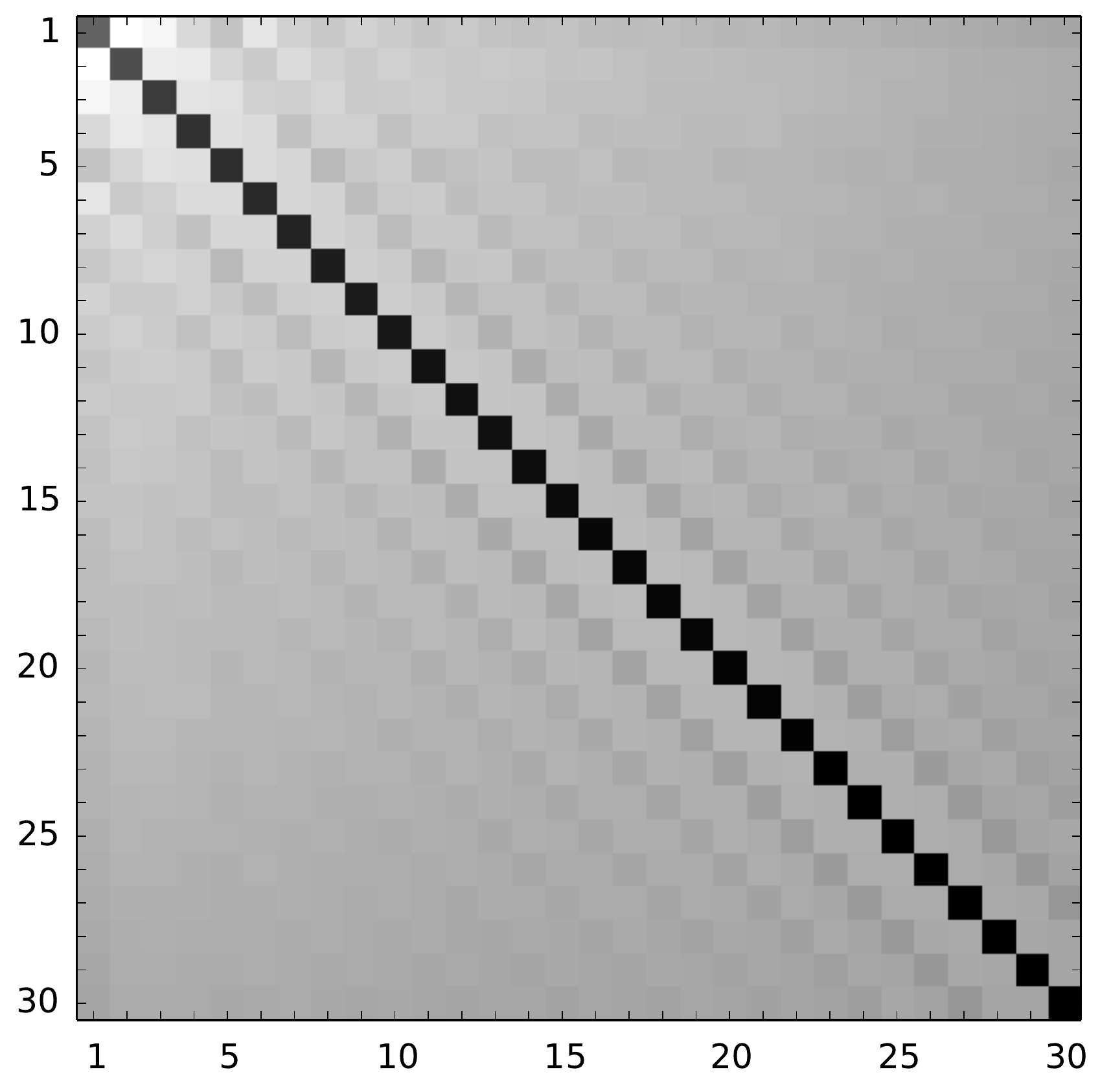}\label{<figure2>}}
     \caption{Average attention weights on positions in the initial list of two PRM models: w/o position embedding.}
     \label{fig:potisions}
\end{figure}

\textbf{Attention on Positions.}
The visualization of average attention weights on different positions in the initial list is shown in Figure~\ref{fig:potisions}. Firstly, Figure~\ref{fig:potisions}(a) showed the self-attention mechanism in our model can capture the mutual influences regardless of the encoding distances as well as the position bias in recommendation list. Items ranked ahead of the list usually is more likely to be clicked and thus have more influences on those items at the tail of the list. For example, we observe that items at the first position have larger impacts on items at 30th position than those items at 26th position even though the latter is more closer to it. The effect of position embeddings is also obvious compared with the Figure~\ref{fig:potisions}(b), whose attention weights between each position are more uniformally distributed.

\section{Conclusion and Future Work}
In this paper, we proposed a personalized re-ranking model (PRM) to refine the initial list given by state-of-the-art learning to rank methods. In the re-ranking model, we used Transformer network to encode both the dependencies among items and the interactions between the user and items. The personalized vector can bring further performance improvements to the re-ranking model. Both the online and offline experiments demonstrated that our PRM model can greatly improve the ranking performance on both public benchmark dataset and our released real-world dataset. Our released real-world dataset can enable researchers to study the ranking/re-ranking algorithms for recommendation systems.

Our work explicitly models the complex item-item relationships in the feature space. We believe that optimization in the label space can also helps. 
Another future direction is learning to diversify by re-ranking. Even though our model does not hurt the ranking diversities in practice. It is worthy to try to introduce the goal of diversification into our re-ranking model. We will further explore this direction in the future work.

\clearpage
\balance
\bibliographystyle{ACM-Reference-Format}


\end{document}